\documentclass[twocolumn,aps,floatfix,superscriptaddress,longbibliography]{revtex4-1}
\pdfoutput=1 
\usepackage{amsmath,amssymb,eucal,graphicx,float,epstopdf,xparse}
\usepackage{epsfig,subfigure}
\usepackage[utf8]{inputenc}

\def \equi#1{\mathrel{\mathop{\kern 0pt\sim}\limits_{#1}}} 
\usepackage[colorlinks=true, urlcolor=blue, anchorcolor=blue, citecolor=blue,filecolor=blue,linkcolor=blue,menucolor=blue]{hyperref}

\begin{document}

\title{Range-controlled random walks}

\author{L. R\'egnier} 
\affiliation{Laboratoire de Physique Th\'eorique de la
  Mati\`ere Condens\'ee, CNRS/Sorbonne Universit\'e, 75005 Paris, France}
\author{O. B\'enichou} 
\affiliation{Laboratoire de Physique Th\'eorique de la
  Mati\`ere Condens\'ee, CNRS/Sorbonne Universit\'e, 75005 Paris, France}
\author{P. L. Krapivsky}
\affiliation{Department of Physics, Boston University, Boston, Massachusetts 02215, USA}
\affiliation{Santa Fe Institute, Santa Fe, New Mexico 87501, USA}

\begin{abstract}
We introduce range-controlled random walks with hopping rates depending on the range $\mathcal{N}$, that is, the total number of previously distinct visited sites. We analyze a one-parameter class of models with a hopping rate $\mathcal{N}^a$ and determine the large time behavior of the average range, as well as its complete distribution in two limit cases. We find that the behavior drastically changes depending on whether the exponent $a$ is smaller, equal, or larger than the critical value, $a_d$, depending only on the spatial dimension $d$. When $a>a_d$, the forager covers the infinite lattice in a finite time. The critical exponent is $a_1=2$ and $a_d=1$ when $d\geq 2$. We also consider the case of two foragers who compete for food, with hopping rates depending on the number of sites each visited before the other. Surprising behaviors occur in 1d where a single walker dominates and finds most of the sites when $a>1$, while for $a<1$, the walkers evenly explore the line. We compute the gain of efficiency in visiting sites by adding one walker.
\end{abstract}

\maketitle

The range $\mathcal{N}(t)$, that is, the number of distinct sites visited at time $t$,  is a central observable of the random walk theory. This quantity has been the subject of a large number of works in various fields, ranging from physics and chemistry to ecology \cite{Weiss-Rubin,Hughes,Weiss,SR}. A key result is that the average range of the symmetric nearest-neighbor random walk exhibits the following asymptotic behaviors~\footnote{We consider hyper-cubic lattices, so e.g. the prediction of \eqref{Nt-12d} in 2d refers to the square grid. }
\begin{equation}
\label{Nt-12d}
 N(t)\equiv\langle \mathcal{N}(t)\rangle\sim
\begin{cases}
\sqrt{\frac{8h t}{\pi}}  & d=1\\
\frac{\pi h t}{\ln h t}       & d=2\\
h t/W_d                      & d>2
\end{cases}
\end{equation}
where $W_d$ are Watson integrals \cite{Watson,GZ_77,G10,Z11} and $h$ the constant hopping rate~\footnote{The hopping rate is $h=zD$ where $z$ is the lattice coordination number and $D$ the diffusion coefficient of the random walker (RW). Hence $h=2d D$ for the RW on the hyper-cubic lattice $\mathbb{Z}^d$.}. The sublinear behavior in $d\leq 2$ dimensions is a direct consequence of the recurrence of random walks in low dimensions.  Beyond the average, the ratio $\sqrt{\text{Var}[\mathcal{N}(t)]}/N(t)$ is known to go to 0 in the large time limit when $d\ge2$; it remains finite for $d=1$. Thus the range $\mathcal{N}(t)$ is asymptotically self-averaging random quantity when $d\geq 2$,  namely its distribution is asymptotically a Dirac delta function peaked at the average value. In 1d, the range is a non-self-averaging random quantity.

In addition to its central  place in random walk theory, the range has proven to be a fundamental tool to quantify the efficiency of random explorations, as it is the case in foraging theory \cite{larralde1992territory,L92,viswanathan1999optimizing,viswanathan2011physics,Havlin2000,benichou2005stochastic,B11}.  The minimal models  involve a forager, described as  a RW, that gradually depletes the resource contained in a medium as it moves. The medium is a $d$-dimensional lattice with a food unit at each site at $t=0$.  When the walker encounters a site containing food, it  consumes it so that the amount of food collected at time $t$ is the range $\mathcal{N}(t)$. This class of models accounts for the depletion of food induced by the motion of the forager, yet the movement of the walker is not affected by the consumption of resources. Depending on the situation, the food collected along the path can provide  additional energy to search for food or, because it represents extra weight, to slow down the walk. As a result, there is  a clear coupling between the range $\mathcal{N}(t)$ and the dynamical properties of the RW. No modeling of this effect has been proposed so far, even at a schematic level. 

\begin{figure}[th!]
    \centering
    \includegraphics[width=\columnwidth]{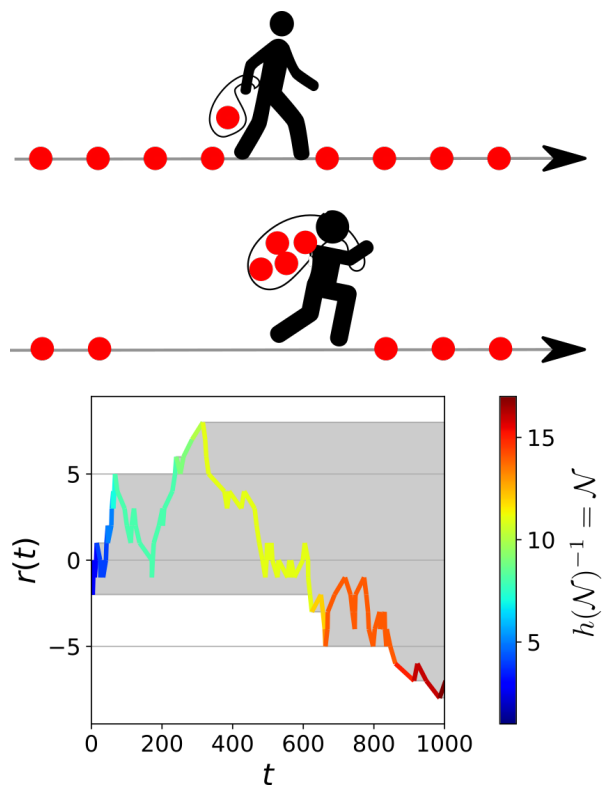}
    \caption{{\bf Representation of a walker penalized upon acquiring new targets.} 
    Top: Representation of the random walk slowed down by the load collected on distinct visited sites. 
    Bottom: Trajectory of a symmetric $1d$ nearest neighbor random walk in the particular case of hopping rates $h(\mathcal{N})$ 
    varying as the inverse of the number of distinct 
    sites visited, $\mathcal{N}^{-1}$. The grey area corresponds to the explored territory, and the coloring refers to the number of distinct sites visited.}
    \label{fig:Example}
\end{figure}

Here, we fill this gap and introduce range-controlled  random walks as a model  accounting for this coupling,  for which the hopping rate is a monotonic function $h(\mathcal{N})$ of the range, either increasing or decreasing (see Fig.~\ref{fig:Example}).  For concreteness, we consider the case where the hopping rate is a power of the amount of collected food: $h(\mathcal{N})=\mathcal{N}^a$. However, our results still apply when this algebraic dependence holds only asymptotically when $\mathcal{N}\gg 1$. In the context of  search problems \cite{B11}, models with positive [resp. negative] exponent $a$ mimic the walker rewarded [resp. penalized] and accelerated [resp. decelerated] upon acquiring new targets (see Fig.~\ref{fig:Example}). Because the coupling between the range and the dynamical properties of foragers is natural and because our modeling of this coupling is minimal, the model of range-controlled random walks quantifies the efficiency of foraging and appears relevant at broader scale to random explorations.

At the theoretical level, range-controlled random walks belong to the class of non-Markovian random walks, in which  the memory of trajectory or some of its features influences the choice of destination sites. Representative examples comprise  self-avoiding walks \cite{RG,MS},  true self-avoiding walkers \cite{Amit,Pietronero,OP83,PP87,G17}, self-interacting random walks \cite{Werner,Davis,volkov,Dickman01,ERW1,ERW2,ERW3,ERW4,Boyer14,Campos19,BV20,BV21} and random walks with reinforcement such as the elephant walks \cite{elephant,elephant-cont,elephant-Gauss,elephant-mass,elephant-Bertoin}. In all of these models, the total hopping rate is kept constant \footnote{Notable exceptions are the locally activated random walk model \cite{BR12}, where the hopping rate depends on the number of visits of a specific site, and the accelerated cover problem \cite{maziya2020dynamically}, where the hopping rate is a function of the typical time to visit a new site.}. Determining the range of non-Markovian random walks is notoriously difficult, and very few exact results are available.

Beyond this theoretical challenge, non-Markovian random walks with memory emerging from the interaction of the walker with the territory already visited are relevant in the case of living cells \cite{mierke2021bidirectional,mierke2010vinculin,mierke2014fundamental,charras2014physical,mierke2019matrix}. It has indeed been observed in vitro \cite{Alessandro2021,flyvbjerg2021}, both in 1d and 2d situations, that various cell types can chemically modify the extracellular matrix, which in turn deeply impact their motility. In this context, range-controlled random walks appears as a minimal model where the modifications induced by the passage of cells are described in a mean-field way: all the complexity of the ``perturbation", be it the concentration field of nutrients \cite{MACKLIN2007}, the local orientation of matrix fibers \cite{SCHLUTER2012} or hydrodynamics fields \cite{HECK2017} is assumed to be encapsulated in the extension of the domain visited by the cell (i.e., the range of the associated random-walk). It is then natural to mimic the ``response" of the cell by a modification of a dynamical parameter, and we finally end up with a hopping rate $h(\mathcal{N})$ depending on the range as introduced above.

{\it Summary of the Results.} In this Letter, we quantify the efficiency  of $d$-dimensional range-controlled random walks by determining  exact asymptotic expressions  of their average range, as well as the full distribution in $d=1$ and in the $d\to\infty$ limit. This allows us to unveil a surprising transition and show that the behavior of range-controlled random walks drastically changes depending on whether the exponent $a$ is smaller, equal, or larger than the critical value, $a_d$, depending only on the spatial dimension: $a_1=2$ in  $d=1$ and $a_d=1$ when $d\geq 2$. The explosive behavior occurs in the supercritical $a>a_d$ regime: The forager covers the entire infinite lattice in a finite time. 

The behavior in the $a\leq a_d$ regime can be appreciated from the growth of the average number $N(t)$ of distinct visited sites. When $a<a_d$, the growth is algebraic with a logarithmic correction in 2d:
\begin{equation}
\label{Nt-a}
N(t)\sim C_d(a)\times
\begin{cases}
t^\frac{1}{2-a}                                        & d=1\\
\left(\frac{t}{\ln t}\right)^\frac{1}{1-a}      & d=2\\
t^\frac{1}{1-a}                                        & d>2
\end{cases}
\end{equation}
 The amplitudes are
\begin{subequations}
 \label{Ca}
 \begin{align}
 \label{Ca:1}
 C_1(a) &= \frac{2^{1+\frac{1}{2-a}}}{\Gamma\left(\frac{1}{2-a}\right)} \int_0^\infty \frac{dv}{v}\left[\frac{\sqrt{v}}{\cosh v}\right]^{\frac{4}{2-a}}, \quad a<2 \\
 \label{Ca:2}
 C_2(a) &= \left[\pi(1-a)^2\right]^\frac{1}{1-a}, \quad a<1 \\
 \label{Ca:d}
 C_d(a) &= \left(\frac{1-a}{W_d}\right)^\frac{1}{1-a}, \qquad a<1,\qquad d\geq 3 \; .
 \end{align}
\end{subequations}
In the critical regime $a=a_d$,  the growth is exponential
\begin{equation}
\label{Nt-crit}
\ln N(t)\sim 
\begin{cases}
\lambda_1 t               & d=1, ~~a=2\\
\lambda_2 \sqrt{t}     & d=2, ~~a=1\\
\lambda_d t               & d>2, ~~a=1
\end{cases}
\end{equation}
with growth rates
\begin{equation}
\label{lambda}
\lambda_1 =1, \quad \lambda_2=\sqrt{2\pi}\,, \quad \lambda_d=1/W_d \quad (d>2)  \; .
\end{equation} 

We discuss the competition between two foragers by determining the average number of distinct sites $N_2(t)$ visited by two foragers in $1d$ where their respective hopping rate depends on the number of distinct sites the walker visited before the other. In particular, by defining $N_1(t)$ as the average number of distinct sites visited by a single walker (without any other walker), we get an analytical value for the ratio at large times:
\begin{equation}
\label{ratio}
\lim_{t\to\infty}\frac{N_2(t)}{N_1(t)} = r_2(a)=\begin{cases}
2^\frac{1-a}{2-a} & a<1 \\
1                          & a>1 \; .
\end{cases}
\end{equation}
This ratio quantifies the efficiency gain in finding new sites by adding one RW. In particular, we observe that for foragers accelerating fast enough with the number of distinct sites visited $(a>1)$, there is no gain in adding the second walker.

{\it 1d.}\label{sec:main-one-d}
Let $P_n(t)\equiv \mathbf{P} (\mathcal{N}(t) = n)$ be the range distribution and $\Pi_n(t)\equiv \mathbf{P} (\mathcal{N}(t) \ge n)$ the corresponding complementary cumulative distribution. In the 1d situation, an exact expression for the entire distribution of the range can be obtained. This exact solution relies on the observation that (see \cite{Leo2022,Regnier2023}) 
\begin{eqnarray}
\lbrace \mathcal{N}(t) \geqslant n \rbrace &=& \left\lbrace \sum\limits_{k=1}^{n-1}\tau_{k} \leqslant t \right\rbrace  
\end{eqnarray}
where $\tau_k$ is the time elapsed between the visit of the $k^\text{th}$ and $(k+1)^\text{st}$ site by the RW defined above. The key points are that (i) during this exploration, the walker has a constant hopping rate, $k^a$, and (ii) the $\tau_k$'s are independent random variables. Performing the Laplace transform  ($\widehat{f}(s)\equiv \int_{0}^{\infty} f(t) e^{-st} dt $) of the probability of these events we get
\begin{eqnarray}
\label{eq:discr-one-d}
   \widehat{\Pi}_n(s)&=&\frac{1}{s}\prod_{k=1}^{n-1} \widehat{F}_{k}(s) 
 \end{eqnarray}
where $\widehat{F}_k$ is the Laplace transform of the distribution of the random variable $\tau_k$. Here $\tau_k$ is the exit time from an interval of $k$ sites starting on the boundary. At small $s$ (corresponding to large time), $\widehat{F}_k$ is given by the exit time distribution of an interval of length $k+1$ starting at distance one of the border of a continuous Brownian motion with diffusion constant $D_k \equiv \frac{1}{2}k^a $ \cite{SR}, 
\begin{eqnarray}
\widehat{F}_k(s)&=&\frac{\sinh\left( \sqrt{\frac{s}{D_k}}k\right)+\sinh\left( \sqrt{\frac{s}{D_k}}\right)}{\sinh\left( \sqrt{\frac{s}{D_k}}(k+1)\right)} \; .
\end{eqnarray}
This expression involves $k/\sqrt{D_k}\propto k^{1-a/2}$, and reveals the existence of three different regimes.

(i) In the subcritical regime $a<a_c=2$, taking the limit $k \to \infty$ and $s \to 0$ while keeping $k^{2-a}s$ finite, gives
\begin{eqnarray}
\label{eq:Fk_one_d}
\widehat{F}_k(s)-1\sim-\sqrt{2s}k^{-a/2}\tanh \left( \sqrt{\frac{s}{2}} k^{1-a/2} \right)
\end{eqnarray}
and then (see Supplementary Material, SM, S1 and S2) 
\begin{align}
  \widehat{P}_n(s) &= \widehat{\Pi}_n(s) - \widehat{\Pi}_{n+1}(s) \nonumber \\
   &\sim -\partial_n \widehat{\Pi}_n(s) \nonumber\\
   & \sim  -\partial_n \left( \frac{1}{s\cosh \left( n^{1-a/2}s^{1/2}/\sqrt{2} \right)^{\frac{4}{2-a}}} \right) \; . 
\label{eq:dist}
\end{align}
In particular, for $a=0$, we recover the well-known range distribution \cite{Hughes} for a standard random walk. One can extract the average range, viz. Eqs.~\eqref{Nt-a} and \eqref{Ca:1}, from the small $s$ asymptotic. In addition, $P_n(t)$ acquires a scaling form $P_n(t)=t^{-\frac{1}{2-a}}\phi_a(x)$ ($x =n/t^{\frac{1}{2-a}}$), where $ \phi_a$ is a function of the scaling variable $x$ depending on the exponent $a$.  Explicit analytical expressions are provided and displayed in SM  for accelerated    and slowed down foragers.

(ii) In the critical regime, $a=a_c=2$ in 1d, $\widehat{F}_k(s)-1\sim-s/k$ and $P_n(t)\sim \delta\left(n-e^t\right)$ (see SM). Thus $\mathcal{N}(t)$ is asymptotically deterministic, chiefly characterized by exponentially growing average $e^{\lambda_1 t}$ with $\lambda_1=1$ as stated in Eq.~\eqref{lambda}.

(iii) In the supercritical regime, $a>a_c=2$, the dynamics is explosive, and the entire infinite lattice is covered in a finite time (see SM).

{\it Higher dimensions.} In the $d\to\infty$ limit, the entire distribution of the range can also be obtained (see SM).  In this case, the average $N$ and the variance $V$ of the number of distinct visited sites exhibit asymptotically identical growth:
\begin{equation}
\label{Nav-inf}
N \sim [(1-a)t]^\frac{1}{1-a} \sim V
\end{equation}
This shows  the self-averaging nature of $\mathcal{N}$ in the sub-critical regime (for $d\to\infty$) as the standard deviation is negligible compared to the average. \\
For finite dimensions, when $a\le a_d$, the asymptotic behavior of the average range can be obtained from heuristic arguments. In the case of varying hopping rates, we use \eqref{Nt-12d} and a self-consistent estimate $h=N^a$ of the typical hopping rate. In 1d, for instance, this leads to $N\propto \sqrt{N^a t}$, from which $N\propto t^\frac{1}{2-a}$, in agreement with the exact solution provided in Eq.~\eqref{Nt-a}. Similarly we arrive at the announced growth laws \eqref{Nt-a} in higher dimensions. These results show that $a_d=1$ when $d\geq 2$. We now turn to the determination of the amplitudes $C_d(a)$ (for $a<a_d$) and the growth rates $\lambda_d$ (for $a=a_d$).

In $d>2$ dimensions, the proper interpretation of \eqref{Nt-12d} is that a RW hops to unvisited sites with probability that approaches  $1/W_d$ \cite{Hughes}. Thus 
\begin{equation}
\label{N-H}
N(t)\sim (W_d)^{-1}H(t)
\end{equation}
where $H$ is the average total number of hops. Using
\begin{equation}
\label{H-av:d}
H(t)\sim \int_0^t d\tau\,[N(\tau)]^a \sim (1-a)[C_d(a)]^a t^\frac{1}{1-a} 
\end{equation}
we arrive at $C_d(a) = (1-a)[C_d(a)]^a/W_d$ leading to the announced result \eqref{Ca:d}. 

\begin{figure}
    \centering
    \includegraphics[width=\columnwidth]{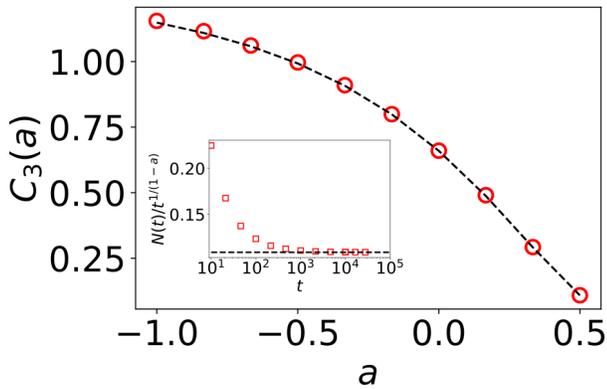}
    \caption{Comparison of the coefficient $C_3(a)$ obtained in \eqref{Ca:d} (black dashed line) and the results of numerical simulations (red circles) obtained by evaluating $\lim\limits_{t \to \infty} \frac{N(t)}{t^{1/(1-a)}}$. The subfigure represents $N(t)/t^{1/(1-a)}$ (red squares) for increasing values of $t$ in the particular case $a=1/2$. It shows the convergence to $C_3(a)$ (black dashed line).}
    \label{fig:C3a}
\end{figure}

The relation \eqref{N-H} is asymptotically exact, but averaging the total number of hops
\begin{equation}
\mathcal{H} = \int_0^t d\tau\,[\mathcal{N}(\tau)]^a
\end{equation}
gives \eqref{H-av:d} only if $\langle \mathcal{N}^a\rangle = \langle \mathcal{N}\rangle^a$. This is erroneous (when $a\ne 1$) if the random quantity $\mathcal{N}$ is non-self-averaging as it is in 1d. Since $\mathcal{N}$ is self-averaging if $d \geq 2$, the prediction \eqref{Ca:d} is exact (see also the agreement with numerical simulations displayed in Fig.~\ref{fig:C3a}). In the critical regime, $a=1$, we have $\dot H=N$ which we insert into \eqref{N-H} and obtain the differential equation 
$\dot H = H/W_d$,
whose solution is $H\propto e^{t/W_d}$. The range is also exponential confirming \eqref{Nt-crit} and \eqref{lambda} for $d>2$. 

In 2d, the exact asymptotic in \eqref{Nt-12d} implies $N\sim \pi H/\ln H$. When $a<1$, taking $h=N^a$ in Eq. \eqref{Nt-12d} in the $2d$ case, we obtain $N\propto \frac{N^a t}{ \ln N^a t} $ and $N\propto \left( \frac{t}{\ln t} \right)^{\frac{1}{1-a}}$. We note the constant prefactor $C_2(a)$, as defined in \eqref{Nt-a}. Then similarly to Eq.~\eqref{H-av:d},
\begin{eqnarray}
\label{H-av:2}
H &\sim& \int_0^t d\tau\,[N(\tau)]^a    \nonumber\\
    &\sim& (1-a)[C_2(a)]^a\, t^\frac{1}{1-a} (\ln t)^{-\frac{a}{1-a}}  \; .
\end{eqnarray}
Equating $N$ to
\begin{equation*}
\frac{\pi H}{\ln H}\sim \pi(1-a)^2[C_2(a)]^a\, (t/\ln t)^\frac{1}{1-a}
\end{equation*}
fixes the amplitude and yields the announced result \eqref{Ca:2}. In the critical $a=1$ regime, the growth is stretched exponential in 2d (see \eqref{Nt-crit}). Indeed, $H(t)$ asymptotically satisfies $\dot H=\frac{\pi H}{\ln H}$,
whose solution is $H\propto  e^{\sqrt{2 \pi t}}$. This confirms \eqref{Nt-crit} with $\lambda_2=\sqrt{2\pi}$ in 2d. Thus we have established \eqref{Nt-crit} and \eqref{lambda} in all dimensions (see SM Fig.~3 for the comparison with numerical simulations).

{\it Two foragers.} We now discuss the competition of foragers. The forager with label $j$ has the hopping rate $\mathcal{F}_j^a$, where $\mathcal{F}_j$ is the number of sites first visited by the forager. For one forager, $\mathcal{F}_1=\mathcal{N}_1$ is just the range. For two foragers, $\mathcal{F}_1+\mathcal{F}_2=\mathcal{N}_2$ is the total range. Foragers do not directly interact, but their motion changes the environment that, in turn, affects the motion of the foragers.  

To compare the two-foragers and the single-forager cases, we consider the ratio $N_2/N_1$ of the average numbers of distinct visited sites in both settings. The ratio $r_2(a)$ defined in Eq.~\eqref{ratio} depends only on the exponent $a$ and is non-trivial only in 1d, as $r_2(a)=2$ for $d\geq 2$ as a consequence of \cite{Paul2022}, the number of common sites visited by the two RWs being asymptotically negligible compared to the number of distinct sites visited by each one of them in this case. Hereinafter we consider foraging in 1d in the non-explosive regime, $a\leq 2$. For 1d RWs, the ratio  $r_2(0)=\sqrt{2}$ is smaller than 2 reflecting the severe space limitation in 1d (see SM).

\begin{figure}
    \centering
    \includegraphics[width=\columnwidth]{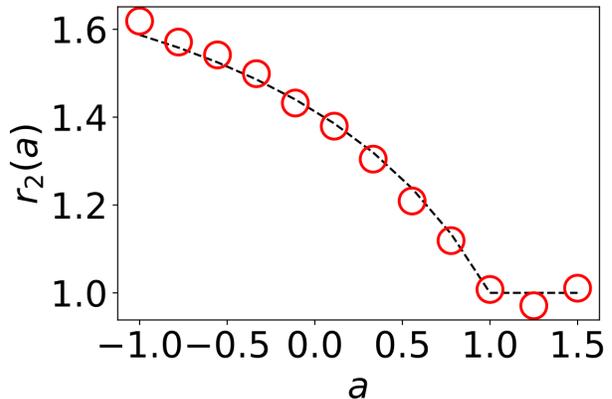}
    \caption{Comparison of the theoretical prediction of the ratio $r_2(a)=\max\left(1,2^{\frac{1-a}{2-a}}\right)$ (black dashed lines) and numerical simulation $\lim\limits_{t\to \infty}\frac{N_2(t)}{N_1(t)}$ (red circles, whose diameters correspond to the largest error bar estimation) of the average number of distinct sites visited by two foragers, $N_2(t)$, versus the average number of distinct sites visited by a single one, $N_1(t)$.}
    \label{fig:2walker}
\end{figure}
 
One has to differentiate between two regimes, $a<1$ and $a>1$. When $a<1$,  both foragers visit the 1d lattice equally (on average). Thus the rate of finding new sites is known by solving the problem at $a=0$. Using a proper rescaling of the times by $(N_2/2)^a$ corresponding to the hopping rate of one of the forager (similarly to what we did with \eqref{Nt-12d}), one establishes \eqref{ratio}. When $a>1$, the hypothesis that sites are equally visited breaks down. To understand the transition, suppose one walker ($W_1$) has found $k\gg 1$ sites, while  he other ($W_2$) has found $\ell\ll k$ sites. If $W_2$ finds a new site at some time $t\gg 1$, which walker will be first to find a new site? The walker $W_2$ will find a new site in a typical time $\propto k/\ell^a$, as it is positioned at the border of the interval of at most $k+\ell \approx k$ distinct sites visited by $W_1$ and $W_2$. The position of $W_1$ is unknown, effectively uniform in the interval of $k$ distinct sites visited, but it hops much faster than $W_2$ and so the average time of finding a new site is $k^2/k^a=k^{2-a}$. Thus, if $a>1$, even though it is further away from the border than $W_2$, the walker $W_1$ will be the first to find a new site ($k^{2-a} \ll k/\ell^a$). This situation is stable if $a>1$ (the dominant walker with the most distinct visited sites  will become more and more dominating) and unstable if $a<1$ (the subdominant walker catches up). The theoretical prediction is validated by numerical simulations (Fig.~\ref{fig:2walker}). 

We introduced random walks with range-dependent hopping rates behaving as  $\mathcal{N}^a$  when $\mathcal{N}\gg 1$. Our analysis provides the exact full distribution of the range  in 1d, and also on a complete graph mimicking an infinite-dimensional setting. For a random walk on a hyper-cubic lattice $\mathbb{Z}^d$ with $d\geq 2$, we used a heuristic approach relying on  results for the classical random walk ($a=0$). We argued that this argument gives asymptotically exact results for the average range when $d\geq 2$. The above sub-critical behaviors occur when $a<a_d$ with $a_1=2$ and $a_d=1$ when $d\geq 2$. When $a>a_d$, the entire infinite lattice $\mathbb{Z}^d$ is covered in a finite time.

%

\end{document}


\RestyleAlgo{ruled}
\title{SUPPLEMENTARY MATERIAL\\
Range-controlled random walks}

\author{L. R\'egnier}
\author{O. B\'enichou}
\author{P. L. Krapivsky}
\maketitle

\tableofcontents

\section{One dimension: Analytical derivations}
\label{sec:one-d}

\begin{figure}
	\centering
	\includegraphics[width=0.5\columnwidth]{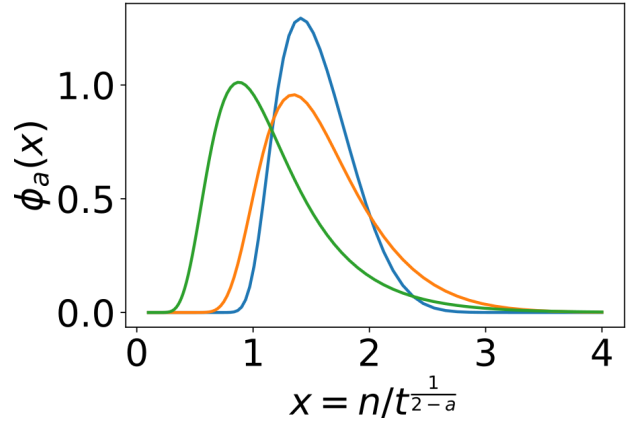}
	\caption{Scaling function $\phi_a(x)$ of the distribution of distinct sites visited for $a=-2$, $a=0$ and $a=1$ 
	(blue, orange and green line respectively) obtained from \eqref{eq:Distrib_am2}, \eqref{eq:Distrib_a0} and \eqref{eq:Distrib_a1} respectively.}
	\label{fig:ScaledDistrib}
\end{figure}

Let us derive the distribution of $\mathcal{N}(t)$, Eq. (11) of the main text. We start from Eq. (8) of the main text, take the logarithm of the product, use the first term in expansion $\ln \widehat{F}_k(s)\sim \widehat{F}_k(s)-1$,  and replace summation by integration to yield
\begin{align}
    \ln \left(\prod_{k=1}^n \widehat{F}_k(s) \right)\sim \int_1^{n} \left(\widehat{F}_k(s)-1 \right) dk 
\end{align}
This is the leading asymptotic contribution when $s \to 0$ and $n\to \infty$. When $a<2$, we integrate Eq. (10)  over $k$ and obtain Eq. (11). 
We derive the expression of $C_1(a)$ given in Eq. (3) from the Tauberian theorem and the $s\to 0$ asymptotic
\begin{align}
    \widehat{N}(s)&\sim \frac{1}{s}\int_0^{\infty}\frac{dn}{\cosh \left( n^{1-a/2}s^{1/2}/\sqrt{2} \right)^{\frac{4}{2-a}}} \nonumber\\
    & \sim \left(\frac{2}{s}\right)^\frac{3-a}{2-a}\int_0^{\infty}\frac{dv}{v}\left(\frac{\sqrt{v}}{\cosh \left( v \right)}\right)^{\frac{4}{2-a}} \; .
\end{align}
For any value of $a$, one can invert Laplace transform (11) to obtain the general scaling form
\begin{align}
	P_n(t)=t^{-\frac{1}{2-a}}\phi_a(x), \qquad x=n/t^{\frac{1}{2-a}}
\end{align}
with scaling functions $\phi_a$ depending on $a$. More explicit results can be obtained for some values of the exponent $a$. The case $a=0$ (no range dependence) is known to be \cite{Hughes}, 
\begin{align}
    \phi_0(x)=\frac{8}{\sqrt{2 \pi }}\sum_{k=1}^\infty (-1)^{k-1}k^2e^{-k^2x^2/2}\; .
    \label{eq:Distrib_a0}
\end{align}
For $a=-2$, one performs the inverse Laplace transform of Eq. (11) and deduces 
\begin{align}
    &\phi_{-2}(x)=\frac{8 \pi}{x^5} \sum_{k=0}^{\infty}(-1)^k (2k+1) e^{-\frac{(2\pi k+\pi)^2}{2x^4}} \; .
    \label{eq:Distrib_am2}
\end{align}
Similarly for $a=1$,
\begin{align}
&\phi_1(x) =\frac{8 }{3 x^5} \sum _{k=0}^{\infty } e^{-\frac{(2 \pi  k+\pi )^2 }{2 x}}   
\Big[(2 \pi  k+\pi)^4    -x^3  -6 \pi ^2 (2 k+1)^2 x  + [(2 \pi  k+\pi )^2+3] x^2 \Big] \; .
\label{eq:Distrib_a1}
\end{align}
In \ref{fig:ScaledDistrib} we display $\phi_a(x)$ for three different values of $a$. 
For $a=2$, $\widehat{F}_k(s)-1\sim-s/k$ results in 
\begin{align}
 \widehat{\Pi}_n(s) = \frac{1}{sn^s} \; .
\end{align}
Performing the inverse Laplace transform we obtain the range distribution $P_n(t)=\delta(n-e^t)$. One can generalize the results corresponding to the algebraic dependence of the hopping rate on the range, $h(\mathcal{N})=\mathcal{N}^a$, to any function of the range. We begin with
\begin{eqnarray}
\widehat{F}_k(s)&=&\frac{\sinh\left( \sqrt{\frac{2s}{h(k)}}k\right)+\sinh\left( \sqrt{\frac{2s}{h(k)}}\right)}{\sinh\left( \sqrt{\frac{2s}{h(k)}}(k+1)\right)}
\end{eqnarray}
and take the $k\to \infty$ and $s\to 0$ keeping $sk^2/h(k)$ fixed; such limit exists only if $h(k)=o(k^2)$. We obtain 
\begin{eqnarray}
\label{eq:Fk_one_d_generalised}
\widehat{F}_k(s)-1\sim-\sqrt{\frac{2s}{h(k)}}\tanh \left( \sqrt{\frac{sk^2}{2h(k)}} \right) \; .
\end{eqnarray}
Using the same method as in the main text we deduce the Laplace transform of $\Pi_n(t)$:
\begin{align}
    \widehat{\Pi}_n(s)=\frac{1}{s}\exp\left[-\int_0^n \sqrt{\frac{2s}{h(k)}} \tanh\left(\sqrt{\frac{sk^2}{2h(k)}} \right) dk \right].
\end{align}
The integral is not easily tractable for an arbitrary function $h$. However, the asymptotic expression given in Eq. (11) still applies for large values of $n$ in the case $h(x)\sim x^a$ as $x\to\infty$ with $a<2$. In this situation, keeping $sn^2/h(n)\sim u=sn^{2-a}$ fixed and taking the $n\to \infty$ and $s\to 0$ limit we obtain 
\begin{align}
    \widehat{\Pi}_n(s) &=\frac{1}{s}\exp\left[-\sqrt{\frac{2sn^2}{h(n)}}\int_0^1 \sqrt{\frac{h(n)}{h(un)}} \tanh\left(\sqrt{\frac{sn^2}{2h(n)}u^2}\sqrt{\frac{h(n)}{h(un)}} \right) du \right] \nonumber \\
    &= \frac{1}{s}\exp\left[-\sqrt{2sn^{2-a}}\int_0^1 \frac{1}{u^{a/2}} \tanh\left(\sqrt{\frac{sn^{2-a}}{2}u^{2-a}} \right) du \right]
\end{align}
which is exactly the primitive of Eq. (11). This further emphasizes the broad range of application of our method: as long as the time rate has a large range algebraic dependence (lesser than quadratic), the distribution in $1d$ at large times is described by Eq. (11).

\section{One dimension: Details and Numerical Verification}
\label{sec:one-d_num}
\begin{figure}[th!]
    \centering
    \includegraphics[width=\columnwidth]{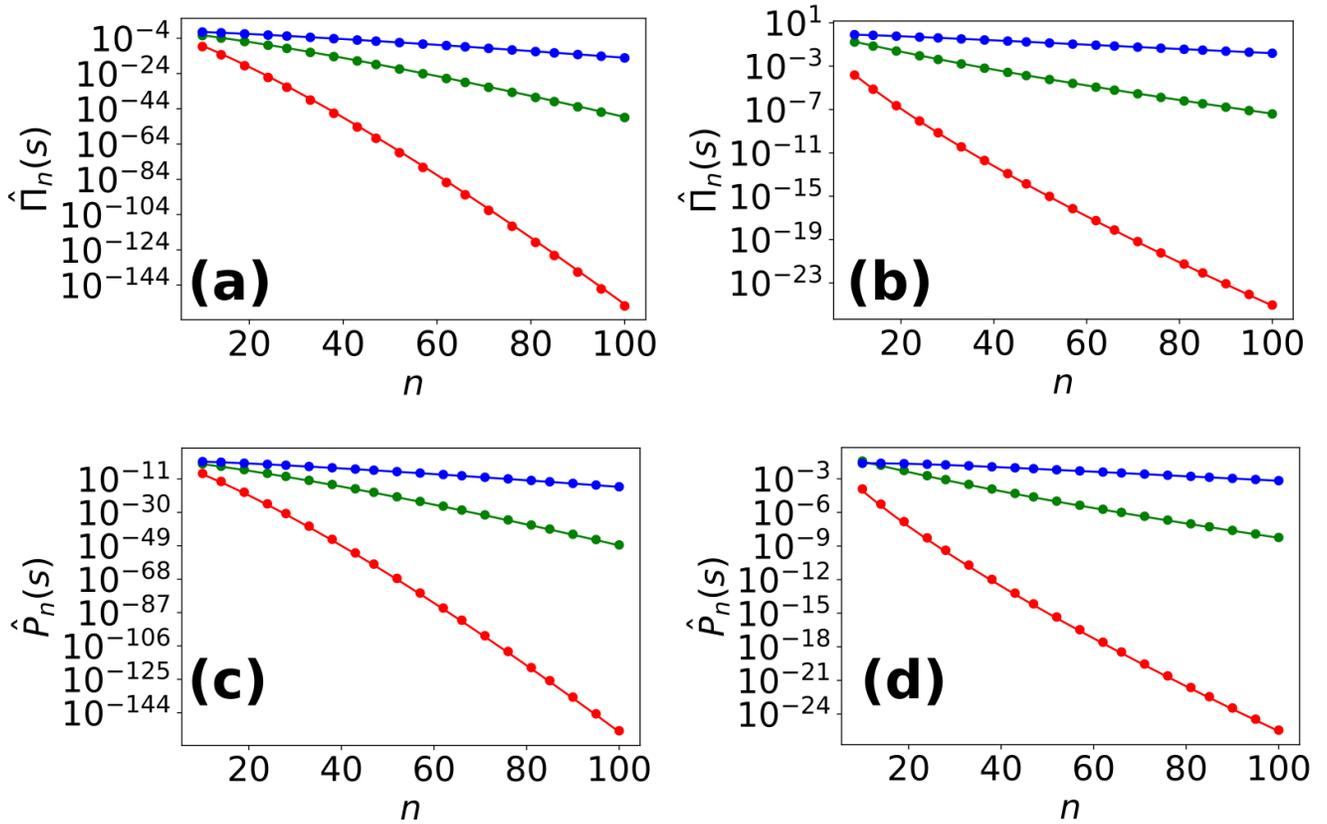}
    \caption{Direct numerical check of the asymptotic relations Eq.~\eqref{eq:Pi} and \eqref{eq:P} for $1d$ calculations. $s$ takes the value $1$, $0.1$ and $0.01$ (in red, green and blue respectively).  \textbf{(a)} and \textbf{(b)} provoide a direct check of \eqref{eq:Pi}, the plain lines standing for the product $\prod_{k=1}^n \hat{F}_k(s)$ while dots stand for $\frac{1}{\cosh \left( n^{1-a/2}s^{1/2}/\sqrt{2} \right)^{\frac{4}{2-a}}}$ $a$ takes the value \textbf{(a)} $-0.5$ and \textbf{(b)} $0.5$. \textbf{(c)} and \textbf{(d)} provide a direct check of \eqref{eq:P}: plain lines standing for the difference $\hat{\Pi}_n(s)-\hat{\Pi}_{n+1}(s)$ while dots stand for the partial derivative $-\partial_n  \hat{\Pi}_n(s)$. $a$ takes the value \textbf{(c)} $-0.5$ and \textbf{(d)} $0.5$. }
    \label{fig:Verif}
\end{figure}
In this section, we check the validity of Eq. (11) of the main text. Our starting point is the exact relation between $P_n(t)=\mathbb{P}(\mathcal{N}(t)=n)$ and $\Pi_n(t)=\mathbb{P}(\mathcal{N}(t)\geq n)$: 
\begin{align}
    P_n(t)&=\Pi_n(t)-\Pi_{n+1}(t) \; .
\end{align}
In the Laplace domain, it becomes the exact relation
\begin{align}
    \hat{P}_n(s)&=\hat{\Pi}_n(s)-\hat{\Pi}_{n+1}(s) \; .
    \label{eq:Pn_and_Pi}
\end{align}
Now, using the expression $\hat{\Pi}_{n+1}(s)=L\lbrace \mathbb{P}(\mathcal{N}(t)>n) \rbrace = \prod_{k=1}^n \hat{F}_k(s)$ (Eq.~(8) in the main text), we obtain asymptotically (large $n$/ small $s$ limit) that
\begin{align}
    \prod_{k=1}^n \hat{F}_k(s)&= \exp \left[ \sum_{k=1}^n \ln \hat{F}_k(s)\right]\\
    &\approx \exp \left[ \sum_{k=1}^n (\hat{F}_k(s)-1)\right] \\
    &\approx \exp \left[ \int_0^n (\hat{F}_k(s)-1) dk \right] \\
    &=\frac{1}{\cosh \left( n^{1-a/2}s^{1/2}/\sqrt{2} \right)^{\frac{4}{2-a}}} \; .
    \label{eq:Pi}
\end{align}
Newt, we use that, asymptotically (large $n$ limit), we have
\begin{align}
    \hat{\Pi}_n(s)-\hat{\Pi}_{n+1}(s) \approx -\partial_n  \hat{\Pi}_n(s) \; .
    \label{eq:P}
\end{align}
Using now \eqref{eq:Pi} and \eqref{eq:P}, we obtain Eq. (11) of the main text, which is asymptotically exact (large $n$/small $s$ limit).

\ref{fig:Verif} provides a direct numerical check of our two asymptotic equalities, and unambiguously confirm the asymptotic validity of Eq. (11) of the main text.

\section{Infinite dimension}
\label{sec:inf-dim}
In the $d\to\infty$ limit, the entire distribution of the range can also be obtained. To mimic the infinite dimension limit, consider a complete graph with $K+1$ vertices. The random walker hops to neighboring sites with equal probability. The probability $P_n(t)$ satisfies
\begin{equation}
\label{PnK}
\dot P_n = \left[1-\tfrac{n-2}{K}\right](n-1)^a P_{n-1}-\left[1-\tfrac{n-1}{K}\right]n^a P_n
\end{equation}
where $n=1,\ldots, K+1$ and $P_n(0)=\delta_{n,1}$. For large graphs, $K\gg 1$, one replaces \eqref{PnK} by an infinite set of linear equations 
\begin{equation}
\label{Pn-inf}
\dot P_n = (n-1)^a P_{n-1} - n^a P_n
\end{equation}
that can be used as long as $n\ll K$. In the critical regime, $a_\infty=1$, we have
\begin{equation}
\label{Pn-1}
\dot P_n = (n-1)P_{n-1} -  n P_n 
\end{equation}
from which 
\begin{equation}
\label{Pn:exp}
P_n = e^{-t} [1- e^{-t}]^{n-1} \; ,
\end{equation}
leading to $N(t)=\sum_{n\geq 1}n P_n(t) = e^t$. The growth rate $\lambda_\infty=1$ corroborates our prediction $\lambda_d=1/W_d$, see Eq.~(5) of the main text, as 
\begin{equation}
\label{Watson}
W_d\equiv\int_0^{2\pi}\ldots\int_0^{2\pi} \left[1-\frac{1}{d}\sum_{i=1}^d \cos q_i\right]^{-1}\prod_{i=1}^d \frac{dq_i}{2\pi} \; 
\end{equation}
goes to $1$ in the $d \to \infty$ limit.\\
In the subcritical regime, $a<1$, we focus on the long time behavior and employ continuum methods, namely we treat $n$ as a continuous variable. The right-hand side of \eqref{Pn-inf} is $Q_{n-1}-Q_n$ with $Q_n = n^a P_n$, so in the continuum approximation it becomes 
 \begin{equation}
 \label{Qmt}
\big(\partial_t + \partial_m)Q=\tfrac{1}{2}\partial_m\left\{[(1-a)m]^{-\frac{a}{1-a}} \partial_m Q\right\}
\end{equation}
with $m=(1-a)^{-1}n^{1-a}$.  Solving \eqref{Qmt} yields 
\begin{equation}
P_n\sim \frac{n^a}{\sqrt{2\pi \tau}}\,\exp\!\left[-\tfrac{(m-t)^2}{2\tau}\right]  , \quad \tau = [(1-a)t]^\frac{1}{1-a} 
\end{equation}
Thus on the complete graph the average $N$ and the variance $V$ of the number of distinct visited sites exhibit asymptotically identical growth:
\begin{equation}
\label{Nav-inf}
N \sim [(1-a)t]^\frac{1}{1-a} \sim V
\end{equation}
This shows  the self-averaging nature of $\mathcal{N}$ in the sub-critical regime (for $d\to\infty$) as the standard deviation is negligible compared to the average.

\section{Finite dimensions}
\label{sec:d-dim}

We derive the asymptotic of the average cover time in the supercritical regime, using that 
\begin{align}
    \label{eq:exact_Tda}
    T_d(a)=\sum_{k=0}^{\infty} \frac{\left\langle \tau_k^d \right \rangle}{k^a}< \infty  \; .
\end{align}
Here $\tau_k^d$ is the time elapsed between the visit of the $k^\text{th}$ and $(k+1)^\text{st}$ site by the random walker of unit hopping rate in dimension $d$ (see \cite{Regnier2023}). The quantity $\left \langle \tau_k^d \right\rangle$ can be interpreted as the hopping rate to unvisited site when the forager has already visited $k$ sites. Thus, from Eq. (22), we obtain (for $a>a_d$) 
\begin{align}
    T_d(a) \propto 
\begin{cases}
\sum\limits_{k=1}^{\infty}\frac{k}{k^a}=\zeta(a-1)          & d=1\\
\sum\limits_{k=1}^{\infty}\frac{\ln(k)}{k^a}=-\zeta ' (a)          & d=2\\
\sum\limits_{k=1}^{\infty}\frac{1}{k^a}=\zeta(a)          & d>2 \; .
\end{cases}
\end{align}
The zeta function has a simple pole, $\zeta(a)\sim\frac{1}{a-1}$ when $a \to 1$. This gives the following asymptotic behaviors close to the critical value,
\begin{align}
\label{Ta:d}
T_d(a) \propto
\begin{cases}
(a-2)^{-1}          & d=1\\
(a-1)^{-2}          & d=2\\
(a-1)^{-1}          & d>2
\end{cases}
\end{align} \\

\begin{figure}[th!]
    \centering
    \includegraphics[width=\columnwidth]{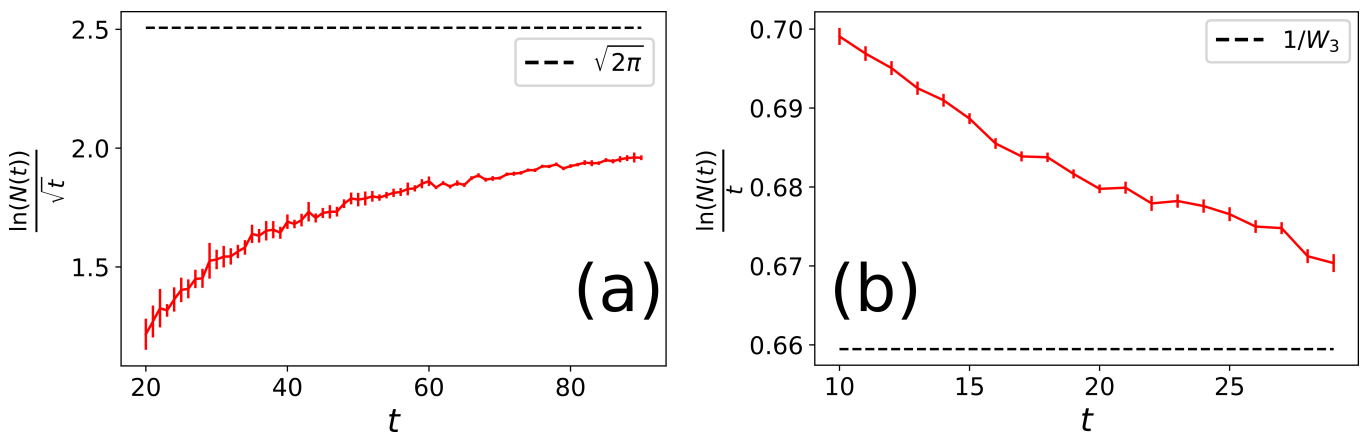}
    \caption{\textbf{(a)} Plot of $\frac{\ln(N(t))}{\sqrt{t}}$ vs time in $d=2$.  \textbf{(b)} Plot of $\frac{\ln(N(t))}{t}$ vs time in $d=3$. 
    Simulation results (red lines) converge to the theoretical predictions for $\lambda_d$, where $\lambda_2=\sqrt{2 \pi}$ and 
    $\lambda_3=1/W_3$ (black dashed lines). The convergence is slow in two dimensions. }
    \label{fig:Critical}
\end{figure}

In \ref{fig:Critical} we plot $N(t)$ at the critical regime, $a=2$, in dimension 2 and 3. Matching simulations results for $N(t)$ with theory is not easy at large times due to the exponential growth, however agreement is quite good. The scaling in $\sqrt{t}$ ($d=2$) and $t$ ($d=3$) for $\ln N(t)$ is particularly convincing; the approach to the predicted growth rate $\lambda_3=1/W_3$ in three dimensions is rather convincing, while in two dimensions the convergence is rather slow. 

\begin{figure}[th!]
    \centering
    \includegraphics[width=\columnwidth]{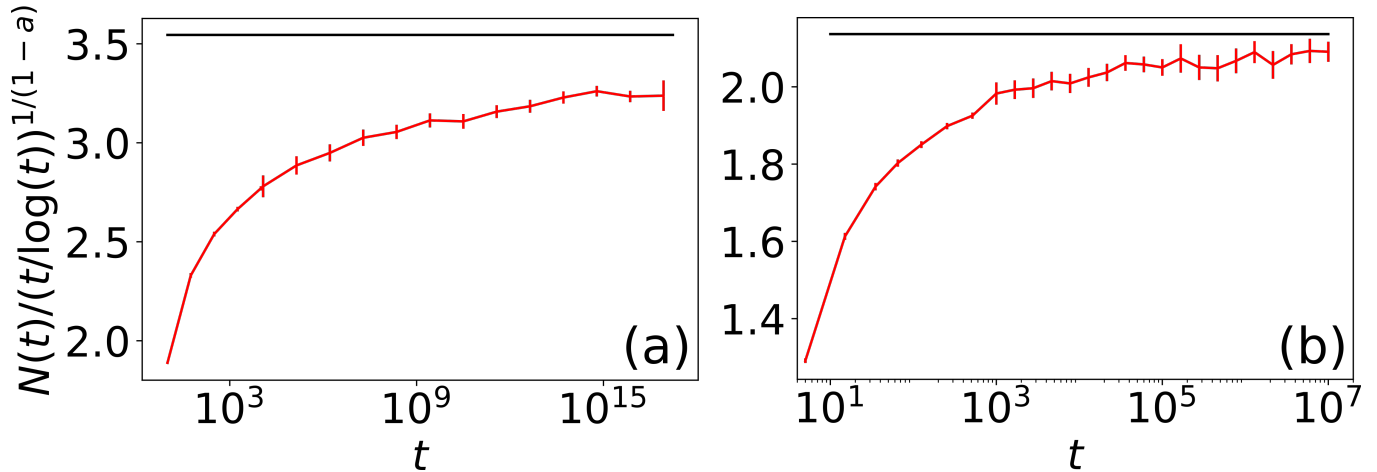}
    \caption{Simulation results $N(t)\left(\frac{\ln(t)}{t}\right)^{-1/(1-a)}$ in two dimensions for \textbf{(a)} $a=-1$ and \textbf{(b)} $a=0.25$. 
    The convergence of simulation results (red lines) to the predicted value [given by Eq.~(3b)] of $C_2(a)$ shown by black lines is 
    significantly slower than in $d=3$, cf. Fig.~2 in the main text. }
    \label{fig:C2a}
\end{figure}
In \ref{fig:C2a}, we verify that the method for obtaining the pre-factor of the dominant term of $N(t)$ for different values of $a$ also works in dimension $d=2$. Despite the fact that the convergence is slow, the limit we obtain from the numerical simulations are in agreement with the theoretical prediction.

\section{Two foragers}
\label{sec:ap-r2}

For random walkers in one dimension, we can employ the continuum approximation. To derive the value of the ratio defined in Eq. (6) at $a=0$ (no range dependence) we use the expression for the probability the Brownian particle never crosses $m$ during the time interval $[0,t]$, 
\begin{align}
S(m,t)\equiv \mathbb{P}[x(\tau)<m|  0\leq \tau\leq t]=\text{Erf}\!\left(\frac{m}{\sqrt{2t}}\right) \; . 
\end{align}
The average number of sites visited by a single random walker during the time interval $(0,t)$ is given by 
\begin{eqnarray}
\label{N1}
N_1& =& 2\int_0^\infty dm\,\{1-S(m,t)\} \nonumber\\
&=& 2\sqrt{2t}\int_0^\infty dz\,\{1- \text{Erfc}(z)\}= \sqrt{8t/\pi}
\end{eqnarray}
where $z=m/\sqrt{2t}$, and in the last step in \eqref{N1} we have used the identity
\begin{equation*}
\sqrt{\pi}\int_0^\infty dz\,\{1- \text{Erf}(z)\} =  1 \; .
\end{equation*}
For two random walkers we similarly obtain
\begin{equation}
\label{N2}
N_2= 2\sqrt{2t}\int_0^\infty dz\,\big\{1- [\text{Erf}(z)]^2\big\}= 4\sqrt{t/\pi}
\end{equation}
where in the last step we have used the identity
\begin{equation*}
\sqrt{\pi}\int_0^\infty dz\,\big\{1- [\text{Erf}(z)]^2\big\} =  \sqrt{2} \; .
\end{equation*}
Equations \eqref{N1} and \eqref{N2} give the announced ratio $r_2(0)=\sqrt{2}$. 

In the single-forager setting we have given heuristic arguments in favor of the rate equation Eq. (22), or
\begin{equation}
\label{N1:eq}
\frac{dN_1}{dt}=\frac{4}{\pi}\,N_1^{-1}N_1^a
\end{equation}
in the present notation. The factor $N_1^a$ on the RHS accounts for the hopping rate, while the factor $\frac{4}{\pi}\,N_1^{-1}$ exhibits the proper scaling with $N_1$ and allows to recover the exact asymptotic answer Eq. (1) for the random walker. 

The analogous equation in the two-foragers setting is
\begin{equation}
\label{N2:eq}
\frac{dN_2}{dt}=\frac{8}{\pi}\,N_2^{-1}(N_2/2)^a
\end{equation}
The factor $(N_2/2)^a$ accounts for the hopping rate and tacitly assumes that both foragers have consumed half food (hypothesis which breaks down for $a > 1$ as explained in the main text). The factor $\frac{8}{\pi}\,N_2^{-1}$ exhibits the proper scaling with $N_2$ and allows to recover the exact asymptotic answer \eqref{N2} for two random walkers. Solving \eqref{N1:eq} and \eqref{N2:eq} yields the announced result Eq. (6) and hence predicts the threshold $a=1$ at which $r_2(a)=1$. \\